\documentclass{article}

\usepackage{arxiv}

\usepackage[utf8]{inputenc} 
\usepackage[T1]{fontenc}    
\usepackage{hyperref}       
\usepackage{url}            
\usepackage{booktabs}       
\usepackage{amsfonts}       
\usepackage{nicefrac}       
\usepackage{microtype}      
\usepackage{lipsum}
\usepackage{graphicx}

\usepackage{cite}
\usepackage{amsmath,amssymb,amsfonts}
\usepackage[linesnumbered,ruled,vlined]{algorithm2e}
\usepackage{textcomp}
\usepackage{bbm}
\usepackage{tabularx}
\usepackage{braket}
\usepackage{xcolor}

\graphicspath{ {./images/} }

\title{Approximate Quantum Circuit Synthesis for Diagonal Unitary}

\author{
Wenqi Zhang, Jinyang Liu, Zixiang Zhou, and Shuai Yang* \\
  School of Cyberspace, Hangzhou Dianzi University\\
  Hangzhou 310018, China \\
  \texttt{*yangshuai24@hdu.edu.cn} \\
}

\begin{document}
\maketitle
\begin{abstract}
The quantum circuit synthesis problem bridges quantum algorithm design and quantum hardware implementation in the Noisy Intermediate-Scale Quantum (NISQ) era. In quantum circuit synthesis problems, diagonal unitary synthesis plays a crucial role due to its fundamental and versatile nature. Meanwhile, experimental results have shown that moderately approximating the original algorithm to conserve quantum resources can improve the fidelity of algorithms during quantum execution. Building on this insight, we propose a quantum circuit synthesis algorithm to design diagonal unitary implementations based on specified quantum resource limits. Our algorithm can synthesize diagonal unitary for quantum circuits with up to 15 qubits on an ordinary laptop. In algorithm efficiency, synthesizing an n-qubit unitary matrix with an exact algorithm requires $2^n$ CNOT gates as a baseline. Within the algorithm error $\varepsilon $ range of interest (0\%–12\%), our algorithm achieves a $3.2\varepsilon $ reduction in CNOT gates on average. In runtime, the algorithm efficiently performs, synthesizing 12-qubit diagonal unitary in an average of 6.57 seconds and 15-qubit in approximately 561.71 seconds.
\end{abstract}

\keywords{Quantum computing \and Quantum circuits \and Diagonal unitary \and Quantum state preparation}

\section{Introduction}
Since its inception, quantum computing has demonstrated remarkable potential in solving classically intractable problems. Some of the most renowned algorithms include the Deutsch-Jozsa algorithm \cite{deutsch1992rapid} for distinguishing balanced and constant functions, Grover’s quantum search algorithm \cite{grover1996fast}, and Shor’s algorithm \cite{Shor1994Polynominal} for large integer factorization. In recent years, quantum machine learning \cite{biamonte2017quantum} and quantum Hamiltonian simulation \cite{berry2015simulating, low2017optimal, low2019hamiltonian, berry2015hamiltonian} have garnered widespread attention due to their superiority over classical algorithms, sparking extensive research and discussion.

Like the trajectory of quantum algorithm development, quantum hardware has seen significant advancements in recent years \cite{kielpinski2002architecture, o2009photonic, barends2014superconducting}. Among these devices, the quantum circuit model has emerged as the most widely utilized framework for quantum computation. Quantum circuits are evaluated across several key resource metrics: circuit size, circuit depth, and the number of ancillary qubits. These resources correspond to the circuit’s final fidelity, execution time, and spatial complexity. In the current NISQ era, however, quantum computing remains constrained by limitations in the number of available qubits, qubit decoherence times, and gate fidelities\cite{preskill2018quantum, arute2019quantum, Wang2016imporved}.

In the current NISQ era, quantum circuit synthesis has undoubtedly emerged as the critical bridge connecting quantum algorithms to quantum hardware. The quantum circuit synthesis problem involves constructing quantum circuits, composed solely of fundamental quantum logic gates, for a given algorithm while minimizing the resources consumed. The concept of quantum circuit synthesis was introduced by Barenco et al. in 1995 \cite{barenco1995elementary}. In this pioneering work, they standardized fundamental quantum logic gates for the first time and proposed the first universal circuit design algorithm for general unitary. Following this, numerous researchers have contributed to the field, leading to many studies on the quantum circuit synthesis problem \cite{knill1995approximation, vartiainen2004efficient, shende2004minimal, mottonen12006decompositions, shende2006synthesis, plesch2011quantum, zhang2021low, sun2023asymptotically, yuan2023optimal, zhang2024parallel}. In these works, researchers identified diagonal unitary as a key issue. Not only can they be easily transformed into state preparation problems, but general unitary can also be decomposed into a series of diagonal unitary. In \cite{sun2023asymptotically}, Sun et al. demonstrated that using phase gadget techniques enables efficient solutions to the design of diagonal unitary across various scenarios. Furthermore, phase gadget, as a special case of Pauli gadget, plays a significant role in quantum chemistry simulation and quantum Hamiltonian simulation.

Given the current limitations in quantum gate fidelities—particularly in circuits composed of multiple quantum gates, where overall fidelity remains low \cite{xie202399}—many researchers have focused on quantum circuit optimization. Their efforts largely target reductions in circuit size \cite{zhou2020monte, huang2022reinforcement, chen2022optimizing, fan2022optimizing} or circuit depth \cite{amy2013meet, li2023single, pointing2024optimizing} to enhance performance. Moreover, recent small-scale quantum computing experiments have revealed that by approximating quantum algorithms to reduce resource consumption (typically by minimizing circuit size), the actual circuit's fidelity can surpass that of an exact algorithm implementation. Consequently, approximate synthesis of quantum circuits has garnered significant attention in recent research \cite{nielsen2002quantum, younis2021qfast, patel2022quest, tan2023quct, gui2024spacetime}.

Based on this, this paper focuses on the synthesis of diagonal unitary. Unlike previous studies, we explore the trade-off between algorithm fidelity and circuit fidelity. Leveraging phase gadget techniques, this work uncovers the significance of phase importance and its superposition properties and designs an approximate synthesis algorithm for diagonal unitary tailored to the NISQ era.

Our contributions in this paper can be summarized as follows:

We conducted an in-depth study of phase gadget techniques, revealing the relationship between phase importance and phase parameters, as well as the superposition property of phase importance when multiple phases are involved. Specifically, when constrained by resource limitations and only able to utilize $k$ phase gadgets, selecting the $k$ phase gadgets with the highest phase importance values is most advantageous for synthesizing diagonal unitary that exhibit minimal error compared to the target unitary.

We designed an approximate algorithm for diagonal unitary tailored to the NISQ era. Through multiple tests using randomly parameterized diagonal unitary, our algorithm achieved favorable results in both algorithm fidelity and runtime. In terms of algorithm fidelity, when an algorithm error $\varepsilon $ of less than 12\%, the algorithm achieves approximately a $3.2\varepsilon $ reduction in the number of CNOT gates. Moreover, the performance of the algorithm improves as the number of qubits increases. Regarding runtime, we measured on a local laptop that when the number of qubits is 12, the algorithm's execution time is approximately 6.57 seconds when the error range is less than 12\%; when the number of qubits is 15, the execution time extends to about 561.71 seconds within the same error range.

\section{Background}
\subsection{Related Works}
The study of circuit synthesis algorithms started in 1995. In \cite{barenco1995elementary}, Barenco et al. first propose a synthesis algorithm to decompose any $n$-qubit unitary matrices, which takes $O(n^3 4^n)$ elementary gates. And then in \cite{shende2006synthesis}, Shende et al. design the synthesis algorithm for general $n$-qubit unitary with only $\frac{23}{48}4^n$ CNOT gates and $O(4^n)$ elementary single-qubit gates. Later, in \cite{plesch2011quantum}, Plesch et al. find the relationship between diagonal unitary synthesis algorithms and general unitary synthesis algorithms. They also give an algorithm that takes $O(2^n)$ elementary gates to synthesize $n$-qubit diagonal unitary. Further, Sun et al. prove that almost all the $n$-qubit diagonal unitary instances require at least $\Omega(2^n)$ elementary gates to synthesize. For the approximating version, in \cite{nielsen2002quantum}, they give the size lower bound for $n$-qubit unitary. With a similar method, we can easily obtain a $\Omega(2^n\log(1/\epsilon)/n)$ size lower bound for $n$-qubit diagonal unitary.  Later in \cite{gui2024spacetime}, they give a synthesis algorithm that can synthesize any $n$-qubit diagonal unitary with $O(2^n\log(1/\epsilon)/n)$ elementary gates.

\subsection{Phase Gadget}
We look at the synthesis problem from the perspective of Boolean algebra. For a diagonal unitary $\Lambda_n=\mbox{diag}(\exp(i\lambda_0),\exp(i\lambda_1),\cdots,\exp(i\lambda_{2^n-1}))$, there is a corresponding Boolean function $f:\{0,1\}^n \to \mathbb{R}$, such that $f(x)=\lambda_x$, and $f$ can be represented by a
real multilinear polynomial. There are $2^n$ terms in the polynomial. For a fixed subset $S\subseteq [n]$, the corresponding monomial is
\[\chi_S(x)=(-1)^{\sum_{i\in S}x_i},\]
and let $\hat{f}(S)$ be the coefﬁcient on monomial $\chi_S(x)$. Then the Fourier expansion of function $f$ is 
\begin{equation}
    f(x) = \sum_{S\subseteq [n]}\hat{f}(S)\chi_S(x).
    \label{equ:Fourierexpansion}
\end{equation}

To synthesize diagonal unitary, we first associate phase gadget with Fourier monomial.
Phase gadgets are an important tool for addressing the design problem of diagonal unitary. A phase gadget is a quantum circuit P that $P_{(s, \alpha)} | x \rangle \to  e^{i \alpha \langle s, x \rangle}$ for any $(s, \alpha) \in \{0, 1\}^n \times \mathbb{R}
$. It consists of two CNOT circuits and an $R_z$ gate. For a Fourier monomial $\chi_S(x)$, the corresponding phase gadget $P_{(s,\hat{f}(S)/2)}$, where $s=\mathbbm{1}(0\in S)\mathbbm{1}(1\in S)\cdots\mathbbm{1}(n\in S)$.

Consider an $n$-qubit diagonal unitary $\Lambda_n = \mbox{diag}(\exp(i\lambda_0),\exp(i\lambda_1),\cdots,\exp(i\lambda_{2^n-1}))$, it can be rewrote by $\mbox{diag}(\exp(if(0)),\exp(if(1)),\cdots,\exp(if(2^n-1)))$. For a phase gadget $P_{s,\hat{f}(S)/2}$, any state $\ket{x}$ transfer to $\exp(\hat{f}(S)\chi_S(x))$, which correspond to a monomial in Fourier expansion.

According to the Fourier expansion of the Boolean function, any $n$-qubit diagonal unitary can be decomposed to $2^n$ monomials, which means $2^n$ phase gadgets. Due to this fact, the synthesis problem for diagonal unitary can be reduced to ``sort'' the phase gadget.

\section{Preliminaries}

\subsection{Problem Definition}
Our goal is to synthesize a diagonal unitary $\Lambda$:
$$
\Lambda = \begin{bmatrix}
\exp(i\lambda_0)  &  &  & \\
  & \exp(i\lambda_1) &  & \\
  &  & ... & \\
  &  &  & \exp(i\lambda_n))
\end{bmatrix}
$$

when the number of qubits is $k$, $\Lambda_k=\mbox{diag}(\exp(i\lambda_0),\exp(i\lambda_1),\cdots,\exp(i\lambda_{2^k-1}))$. We denote our target diagonal unitary as $U_T$, which is determined by the real parameter vector $\lambda _{T} = (\lambda _{1}, \lambda _{2}, ..., \lambda _{2^k - 1})$.  $\lambda _{T}$ is derived from the real parameter vector $\alpha _{T} = (\alpha _{1}, \alpha _{2}, ..., \alpha _{2^k - 1})$ through the Walsh-Hadamard transformation, where the parameters in vector $\alpha _{T}$ represent the rotation angles of the $R_z$ gates within the phase gadgets. The task is choosing some proper phase gadgets and listing them in optimized order. W.L.O.G, let the target qubit of different phase gadgets be the same one in this manuscript.


\subsection{Algorithm Evaluation Metric}
The quality of the algorithm is assessed based on the circuit length it produces and the fidelity achieved. The circuit length provides an optimality criterion for the synthesis algorithm, with shorter circuits being preferable. Given the lower fidelity of CNOT gates on NISQ devices, the number of CNOT gates becomes a direct metric for measuring the overall circuit length; therefore, we aim to minimize this quantity. The algorithm fidelity is determined by the error between the diagonal unitary $U_C$ synthesized by the algorithm and the target diagonal unitary $U_T$, and we denote our error as $D(U_T, U_C)$.

When determining the measurement error $D(U_T, U_C)$ of the solution, it is essential to select an appropriate metric to accurately reflect the differences between unitary. The choice of metric is guided by the principle that the measurement value for identical unitary is zero, while the measurement value increases for unitary that exhibit greater differences, with an error value not exceeding 1. In measuring the distance between unitary, many scholars have used the rank of the product of the original matrix and the conjugate transpose of the target, such as $1 - {\mbox{Re}(\mbox{Tr}(U_{T}^{\dagger } U_C))}/{n}$ \cite{younis2021qfast} and $\sqrt{1 - {{\parallel \mbox{Tr}(U_{T}^{\dagger } U_C) \parallel }^2}/{n^2}} $ \cite{patel2022quest}. However, for a diagonal unitary, the off-diagonal elements are all zero; thus, we only need to consider the diagonal elements. Consequently, we utilize the 2-norm to measure the error:
\begin{equation}
    D(U_C, U_T) = \left( \sum_i \left( \frac{e^{i\lambda} - e^{i\lambda'}}{2} \right)^2 \right)^{1/2}
\end{equation}

The algorithm utility is defined by the ratio of the proportion of CNOT gates saved to the error $D(U_T, U_C)$. A higher algorithm utility ratio is achieved either by increasing the proportion of CNOT gates saved or by reducing the error between the synthesized diagonal unitary and the target unitary. This ratio serves as an excellent metric for evaluating the algorithm.

\subsection{Importance of Phases and Their Superposition Properties}
We have identified the relationship between the error $D(U_C, U_T)$ and the rotation angles of the $R_z$ gates in the phase gadgets (vector $\alpha_T$), which we refer to as phase importance. We also discovered the superposition properties of phase importance. These findings are of significant relevance for our subsequent work on the ordering of phase gadgets.

\subsubsection{Phase Importance}
Regarding the importance of the phase, we have observed that the larger the rotation angle of the $R_z$ gate in a phase gadget, the greater its importance. Given limited resources, we assume that only $2^k-1$ phase gadgets can be utilized, meaning one phase gadget from the complete set must be excluded. Experimental results indicate that when a phase gadget with a larger $R_z$ rotation angle is discarded, the overall error increases significantly, suggesting that such a phase is more critical and should be prioritized. Conversely, when the $R_z$ rotation angle is closer to zero, the corresponding phase gadget results in a minimal error, implying its lesser importance and a lower priority for selection. As shown in Table \ref{Table of Error Incurred by Discarding a Phase and Its Corresponding Rz Gate Rotation Angle in this Phase}, for qubits is 5, the relationship between the error incurred by discarding a phase gadget and the rotation angle of the Rz gate is evident. Hence, we assign a importance value to each phase based on the rotation angle of the $R_z$ gate in the respective phase gadget.

\begin{table}[htbp]
    \centering
    \caption{Table of error incurred by discarding a phase and its corresponding Rz gate rotation angle in this phase}
    \label{Table of Error Incurred by Discarding a Phase and Its Corresponding Rz Gate Rotation Angle in this Phase}
    \begin{tabular}{p{1cm} p{1.53cm} p{1cm} p{1cm} p{1.53cm} p{1cm} p{1cm} p{1.53cm} p{1cm}}
        \toprule
        Index & $\alpha_T$ & Error & Index & $\alpha_T$ & Error & Index & $\alpha_T$ & Error\\
        \midrule
        21 & 0.0175  & 0.0087  & 23 & -0.1376  & 0.0688  & 4 & 0.3480  & 0.1731   \\ 
        8 & -0.0296  & 0.0148  & 18 & 0.1487  & 0.0743  & 6 & 0.3844  & 0.1910   \\ 
        17 & 0.0409  & 0.0205  & 9 & -0.1648  & 0.0823  & 7 & -0.3860  & 0.1918   \\ 
        25 & -0.0478  & 0.0239  & 12 & 0.1719  & 0.0859  & 29 & -0.3891  & 0.1933   \\ 
        5 & 0.0555  & 0.0277  & 31 & -0.2298  & 0.1147  & 27 & 0.4631  & 0.2295   \\ 
        14 & -0.0628  & 0.0314  & 22 & 0.2431  & 0.1213  & 3 & -0.4804  & 0.2379   \\ 
        30 & 0.0752  & 0.0376  & 26 & 0.2774  & 0.1383  & 16 & 0.5047  & 0.2497   \\ 
        10 & 0.0958  & 0.0479  & 15 & -0.3147  & 0.1567  & 2 & -0.5961  & 0.2937   \\ 
        28 & -0.1144  & 0.0572  & 1 & 0.3161  & 0.1574  & 13 & -0.9482  & 0.4565   \\ 
        19 & 0.1174  & 0.0587  & 11 & -0.3258  & 0.1622  & 0 & 2.7625  & 0.9821   \\ 
        24 & -0.1365  & 0.0682  & 20 & -0.3464  & 0.1723  & ~ & ~ &   \\ 
        \bottomrule
    \end{tabular}
\end{table}

\subsubsection{Superposition Properties of Phase Importance}  
Given that each phase gadget possesses an importance value, we hypothesize that when selecting $k$ phase gadgets, the set that minimizes the overall error would consist of the $k$ phase gadgets with the highest importance values. To test this hypothesis, we experimented with qubits is 5, where only $2^k-4$ phase gadgets could be chosen, meaning four gadgets had to be excluded from the universal set. We then calculated the relationship between the error and the indices of the discarded phase gadgets. Table \ref{Table of the Relationship Between Error and Discarded Phases} shows that the lowest error was obtained when the four phase gadgets with the smallest importance values were used. The second lowest error occurred when the three phase gadgets with the smallest importance values and the fifth phase gadget were used. This experiment supports our hypothesis and provides valuable insights into the optimal selection of phase gadgets for future implementations.

\begin{table}[htbp]
    \centering
    \caption{Table of the relationship between error and discarded phases}
    \label{Table of the Relationship Between Error and Discarded Phases}
    \begin{tabular}{p{1cm} p{1cm} p{1cm} p{1cm} p{1cm} p{1cm} p{1cm} p{1cm} p{1cm} p{1cm}}
        \toprule
        Error & phase & phase & phase & phase & Error & phase & phase & phase & phase \\
        \midrule
0.0358~ & 8  & 17 & 21 & 25 & 0.0468~ & 8  & 14 & 17 & 25  \\
0.0385~ & 5  & 8  & 17 & 21 & 0.0474~ & 5  & 14 & 17 & 21  \\
0.0404~ & 5  & 8  & 21 & 25 & 0.0477~ & 8  & 21 & 25 & 30  \\
0.0412~ & 8  & 14 & 17 & 21 & 0.0489~ & 5  & 8  & 14 & 17  \\
0.0428~ & 5  & 17 & 21 & 25 & 0.0490~ & 5  & 14 & 21 & 25  \\
0.0430~ & 8  & 14 & 21 & 25 & 0.0498~ & 5  & 8  & 21 & 30  \\
0.0445~ & 5  & 8  & 17 & 25 & 0.0498~ & 17 & 21 & 25 & 30  \\
0.0453~ & 5  & 8  & 14 & 21 & 0.0504~ & 5  & 8  & 14 & 25  \\
0.0453~ & 14 & 17 & 21 & 25 & 0.0512~ & 8  & 17 & 25 & 30  \\
0.0461~ & 8  & 17 & 21 & 30 & 0.0517~ & 5  & 17 & 21 & 30 
 \\
        \bottomrule
    \end{tabular}
\end{table}

\section{Diagonal Unitary Synthesis}
In this section, we will describe the synthesis algorithm for diagonal unitary. We decompose the diagonal unitary into several phase gadgets and carefully reorder these phase gadgets to minimize the CNOT cost while maximizing algorithm fidelity. This section fully leverages the capabilities of phase gadgets. Since one key advantage of phase gadgets is the ability to optimize the CNOT gates between them when two gadgets are combined, an important concept is to use a limited number of CNOT gates to achieve a greater number of phases, prioritizing those with higher importance. Therefore, we determine the number of CNOT gates here to minimize the error $D(U_T, U_C)$ of the diagonal unitary. We will present this in two parts: the problem of phase gadget ordering and the circuit synthesis algorithm.

\subsection{Phase Gadget Ordering Problem}
In this section, we transform our phase gadget ordering problem into a path search problem on a graph. We will construct an undirected graph where each phase $s$ represents a node, and the phase importance serves as the weight of the nodes. Since we are focusing on the synthesis of diagonal unitary on a complete graph, we connect the nodes of the two phases if the number of CNOT gates between them can be reduced to a single CNOT gate. Starting from node $s=0^k$, we traverse a path where the order of the nodes along this path yields the phase gadget ordering results.

We will divide our search algorithm into five components: phase importance calculation, path selection, path extension strategy, ``dead-end'' handling, and path search.

\subsubsection{Phase Importance Calculation}
In this section, we calculate and quantify phase importance, as illustrated in Algorithm \ref{Phase Importance}. Additionally, we apply a logistic function to the computed results to amplify the differences in phase importance near the critical point.

\begin{algorithm}[!h]
\setlength{\algomargin}{1em} 
\SetKwData{Left}{left}\SetKwData{This}{this}\SetKwData{Up}{up} \SetKwFunction{Union}{Union}\SetKwFunction{FindCompress}{FindCompress} \SetKwInOut{Input}{input}\SetKwInOut{Output}{output}

    \KwData{the parameter vector $\alpha _{T}$, the number of qubits $k$, the number of CNOT gates $C$, an adjustment parameter $\gamma$}
    \KwResult{the phase importance $Imp$}
 
    \BlankLine 
	 
    \emph{$Imp = [0] * 2^k$}
    
    \For{each $i \in [1, 2^k-1]$}{
        $Imp[i]$ = $abs(\alpha _{T}[i])$
    }
    
    \emph{$Imp = standardization_{01}(Imp)$}
    
    \emph{$Imp[0] = 1$}
    
    \emph{$temp = FindKthLargest(\alpha _{T}, C + 2)$} 
    
    \For{each $i \in [1, 2^k-1]$}{
        \emph{$Imp^{adjusted} = (Imp[i] - temp) * \gamma$}
        \emph{$Imp[i] = \frac{1}{1 + e^{-Imp^{adjusted}} } $}
    }

    \caption{Phase Importance Calculation}
    \label{Phase Importance}
\end{algorithm}

\subsubsection{Path Selection}
In this section, we establish criteria for selecting future nodes along the search path, as shown in Algorithm \ref{Path Selection}. First, for the tail node $v_k$ in path $u=v_1, v_2,..., v_k$, we count the number of active neighbor nodes. If there are no active nodes among the neighbors, we exit the process; otherwise, we select the unvisited neighbor node with the highest importance value as the next node $v_{k+1}$ in the path. We then repeat this process with node $v_{k+1}$ as the new tail node $v_{k}$ until we exit. Here, an active node is defined as one that has not appeared in the path and has an importance value greater than $0.5-\varepsilon $. The parameter $\varepsilon $ represents the importance relaxation value, which is used to account for phase nodes whose importance values are at the critical threshold.

\begin{algorithm}[!h]
\setlength{\algomargin}{5em} 
\SetKwData{Left}{left}\SetKwData{This}{this}\SetKwData{Up}{up} \SetKwFunction{Union}{Union}\SetKwFunction{FindCompress}{FindCompress} \SetKwInOut{Input}{input}\SetKwInOut{Output}{output}
    
    \KwData{the path $path$, the phase importance $Imp$, the number of CNOT gates $C$, the relaxation value $\varepsilon$}
    \KwResult{a new path $path$}
 
    \BlankLine 

    \While{$len(path) < C+2$}{
        \emph{$neighbor^{active} = []$}
        
        \For{each $node \in path[-1].neighborNode$}{
            $Imp[i]$ = $abs(\alpha _{T}[i])$
            
            \If{$node.Imp > 0.5 - \varepsilon$ and $node \notin path$}{
                $neighbor^{active} \gets neighbor^{active} + node$
            }
            \If{$len(neighbor^{active}) == 0$}{
                break
            }
            \Else{
                $node^{next} = MaxImp(neighbor^{active})$
                $path \gets path + node^{next}$
            }
        }
    }

    \caption{Path Selection}
    \label{Path Selection}
\end{algorithm}

\subsubsection{Path Extension Strategy}
In this section, we introduce the path extension strategy determined when the path selection is obstructed and needs to exit, as illustrated in Algorithm \ref{Path Extension}. Inspired by the path extension strategies derived from Hamiltonian path judgments \cite{dharwadker2004new}, we establish this extension strategy. For path $u=v_1, v_2,..., v_k$, if node $v_i$ is a neighbor of node $v_k$ and there exists a neighbor node $v_{i+1}$ with an active neighbor node $v_j$, a new path $u=v_1, v_2,..., v_i, v_k, v_{k-1},..., v_{i+1},v_j$ is generated. If no such node $v_i$ exists, we exit the process.

\begin{algorithm}[!h] \SetKwData{Left}{left}\SetKwData{This}{this}\SetKwData{Up}{up} \SetKwFunction{Union}{Union}\SetKwFunction{FindCompress}{FindCompress} \SetKwInOut{Input}{input}\SetKwInOut{Output}{output}
    
    \KwData{the path $path$, the phase importance $Imp$, the relaxation value $\varepsilon$}
    \KwResult{a new path $path$}
    
    \emph{$tailNode = path[-1]$ }
    
    \emph{$MaxImp$ = 0}
    
    \emph{$NextNode$ = None} 
    
    \emph{$NextNodeI$ = -1}

    \For{each $node \in tailNode.neighborNode$}{
        \If {$node \in path[:-2]$}{
            $index$ = $path$.index($node$)

            \For{each $node \in path[index + 1].neighborNode$}{
                \If {$node \notin path$ and $node.Imp > MaxImp$}{
                    $NextNode$ = $node$

                    $NextNodeI$ = $index$
                }
            }
        }
    
    }

    \If{$NextNode.Imp > 0.5 - \varepsilon$}{
        $path = path[:index] + path[index + 1:][::-1] + [NextNode]$
    }

    \caption{Path Extension}
    \label{Path Extension}
\end{algorithm}

\subsubsection{``Dead-end'' Handling}
We refer to the situation where both path selection and path extension strategies become obstructed and exit as a ``dead-end''. This section addresses future node selection in such cases, as illustrated in Algorithm \ref{``dead-end'' Handling}. 

\begin{algorithm}[!h] \SetKwData{Left}{left}\SetKwData{This}{this}\SetKwData{Up}{up} \SetKwFunction{Union}{Union}\SetKwFunction{FindCompress}{FindCompress} \SetKwInOut{Input}{input}\SetKwInOut{Output}{output}
    
    \KwData{the path $path$, the phase importance $Imp$, the weight coefficient $\omega$}
    \KwResult{a new path $path$}
    
    \emph{$maxScore$ = 0}
    
    \emph{$nextNode$ = None}

    \For{each $node_i \in tailNode.neighborNode$}{
        \If {$node_i \notin path$}{
            $score = score + node_i.Imp$
        }

        \For{$node_j \in allNode$}{
            \If {$node_j \notin path$ and $node_i \neq node_j$}{
                $score += \omega \cdot (n - d(node_i \to  node_j))\cdot node_j.Imp$
            }
        }
        \If {$score > maxScore$}{
            $nextNode = node_i$
        }
    }
    
    \emph{$path = path + nextNode$}

    \caption{``Dead-end'' Handling}
    \label{``dead-end'' Handling}
\end{algorithm}
When the path cannot be extended at an active node, we consider alternative nodes. Here, we assess the neighbor nodes of the tail node, taking into account both the phase importance of the nodes and the distance score relative to other active nodes:
\begin{equation}
    Score_{i} = Imp_{i} + \omega \sum_{i\neq j, j \notin path} ((n-d(i \to  j))Imp_{j})
\end{equation}
where $\omega$ is the weight coefficient between the two parts, and $d(i \to j)$ is the distance between the two nodes. When evaluating the phase importance of a node, a higher importance value corresponds to a higher score; if the node has already appeared in the path, its score is set to zero. For the distance evaluation between the node and other active nodes, the score increases as the node gets closer to those with higher importance values. The underlying idea of this section is that, although we cannot reach the active nodes from this node, we select a node that may facilitate our future access to the active nodes.

\subsubsection{Path Search}
In this section, we integrate the previous modules to form a complete path search algorithm. Additionally, we consider the relaxation value $\varepsilon$ for the phase importance when defining active nodes, as detailed in Algorithm \ref{Path Search}.

\begin{algorithm}[!h] \SetKwData{Left}{left}\SetKwData{This}{this}\SetKwData{Up}{up} \SetKwFunction{Union}{Union}\SetKwFunction{FindCompress}{FindCompress} \SetKwInOut{Input}{input}\SetKwInOut{Output}{output}
    
    \KwData{the parameter vector $\lambda _{T}$, the number of qubits $k$, the number of CNOT gates $C$, an adjustment parameter $\gamma$}
    \KwResult{a path $path$}
    
    \emph{$\alpha _{T} = \frac{1}{2^k} H^{\oplus k}\lambda _{T}$}
    
    \emph{$Imp = PhaseImportance(\alpha _{T}, k, C, \gamma)$}
    
    \emph{$path = [0^k]$}
    
    \emph{$path = PathSelection(path, Imp, C, 0)$}
    
    \For{each $\omega \in [0.01: 0.5]$}{
        \If {$len(path) \ge C+2$}{
            break
        }

        \While{$len(path) < C+2$}{
            $path = PathExtension(path, Imp, \omega)$
            
            $path = PathSelection(path, Imp, C, \omega)$
        }
    
    \emph{$path = DeadEndHandling(path, Imp, \omega)$}
    }

    \caption{Path Search}
    \label{Path Search}
\end{algorithm}

\subsection{Circuit Synthesis Algorithm}
In this section, based on the sorted phase gadgets and the parameters vector $\alpha_T$, we can determine the real parameter vector $\alpha_C$. By applying the Walsh-Hadamard transformation, we can obtain the diagonal unitary synthesized by our algorithm, as illustrated in Algorithm \ref{Circuit Synthesis}.

\begin{algorithm}[!h] \SetKwData{Left}{left}\SetKwData{This}{this}\SetKwData{Up}{up} \SetKwFunction{Union}{Union}\SetKwFunction{FindCompress}{FindCompress} \SetKwInOut{Input}{input}\SetKwInOut{Output}{output}
    
    \KwData{the parameter vector $\lambda _{T}$, the number of qubits $k$, the number of CNOT gates $C$, an adjustment parameter $\gamma$}
    \KwResult{The quantum circuit $Circuit$}
    
    \emph{$path = PathSearch(\lambda _{T}, k, C, \gamma)$}
    
    \emph{Push the gate $X(x_i)$ into $Circuit$}
    
    \emph{Push the gate $R_z(\alpha_{T}[path[1].phase])$ into $Circuit$}
    
    \For{$node \in path[2:]$}{
        Push the gate $CNOT(x_i, x_j)$ into $Circuit$ 
        
        Push the gate $R_z(\alpha_{T}[node.phase])$ into $Circuit$ 
    }
    \caption{Circuit Synthesis}
    \label{Circuit Synthesis}
\end{algorithm}

\section{Experiments}
We tested the performance of our algorithm under different numbers of qubits, with the parameter vectors $\lambda_T$ for the target diagonal unitary $U_T$ being randomly generated. Both error $D(U_T, U_C)$ and runtime were averaged over multiple test runs. Our experimental results are presented in Table \ref{The error values and runtimes}.

\begin{table*}[h!]
\centering
\caption{The error values and runtimes obtained under different qubit counts and various CNOT gate reduction ratios}
\label{The error values and runtimes}
\begin{tabularx}{\textwidth}{X X X X X X X X X X X X}
\toprule
\multicolumn{3}{c}{qubits=8} & \multicolumn{3}{c}{qubits=10} & \multicolumn{3}{c}{qubits=12}  & \multicolumn{3}{c}{qubits=15} \\
\cmidrule(lr){1-3} \cmidrule(lr){4-6} \cmidrule(lr){7-9} \cmidrule(lr){10-12}
ReCNOT & Error & Time(s) & ReCNOT & Error & Time(s) & ReCNOT & Error & Time(s) & ReCNOT & Error & Time(s) \\
\midrule
50\% & 0.2677 & 0.0359 & 50\% & 0.2560 & 0.4879 & 50\% & 0.2508 & 8.3920 & 50\% & 0.2470 & 840.1679 \\
45\% & 0.2288 & 0.0337 & 45\% & 0.2192 & 0.4282 & 45\% & 0.2154 & 7.2275 & 45\% & 0.2106 & 627.0197 \\
40\% & 0.1923 & 0.0331 & 40\% & 0.1842 & 0.3984 & 40\% & 0.1810 & 7.0614 & 40\% & 0.1763 & 642.0624 \\
35\% & 0.1572 & 0.0308 & 35\% & 0.1524 & 0.3804 & 35\% & 0.1468 & 6.6568 & 35\% & 0.1437 & 584.1792 \\
30\% & 0.1266 & 0.0314 & 30\% & 0.1199 & 0.3787 & 30\% & 0.1167 & 6.6093 & 30\% & 0.1146 & 580.7165 \\
25\% & 0.0986 & 0.0290 & 25\% & 0.0934 & 0.3696 & 25\% & 0.0907 & 6.4527 & 25\% & 0.0878 & 552.9863 \\
20\% & 0.0749 & 0.0304 & 20\% & 0.0686 & 0.3819 & 20\% & 0.0661 & 6.4664 & 20\% & 0.0642 & 550.4549 \\
15\% & 0.0526 & 0.0275 & 15\% & 0.0475 & 0.3909 & 15\% & 0.0458 & 6.6140 & 15\% & 0.0437 & 564.0758 \\
10\% & 0.0348 & 0.0293 & 10\% & 0.0316 & 0.3925 & 10\% & 0.0284 & 6.6088 & 10\% & 0.0263 & 557.5657 \\
5\% & 0.0217 & 0.0286 & 5\% & 0.0179 & 0.3871 & 5\% & 0.0154 & 6.6715 & 5\% & 0.0130 & 564.4382 \\
\bottomrule
\end{tabularx}
\end{table*}

\subsection{Experimental Setup}
We implemented our algorithm in Python 3.12 and NumPy 2.0.0. The timing tests were conducted on a local laptop equipped with a 2.3 GHz Intel(R) Core(TM) i7-12700H CPU.

\subsection{Algorithm Utility Ratio}Within the error range of interest (0\%–12\%), our algorithm utility ratio can reach between 2.5 and 3.85, with an average algorithm utility ratio of 3.2. In the case of qubits is 15, we can save 30\% of the CNOT gates at the cost of an algorithm fidelity loss of 8.78\%; at the same qubit count, we can save 15\% of the CNOT gates at the cost of an algorithm fidelity loss of 4.37\%; and we can save 5\% of the CNOT gates at the cost of an algorithm fidelity loss of 1.30\%. Therefore, our algorithm is significant in terms of cost savings and improving circuit fidelity. Additionally, we have identified the following two characteristics.

\begin{figure}[!h]
\centering
\includegraphics[scale=0.35]{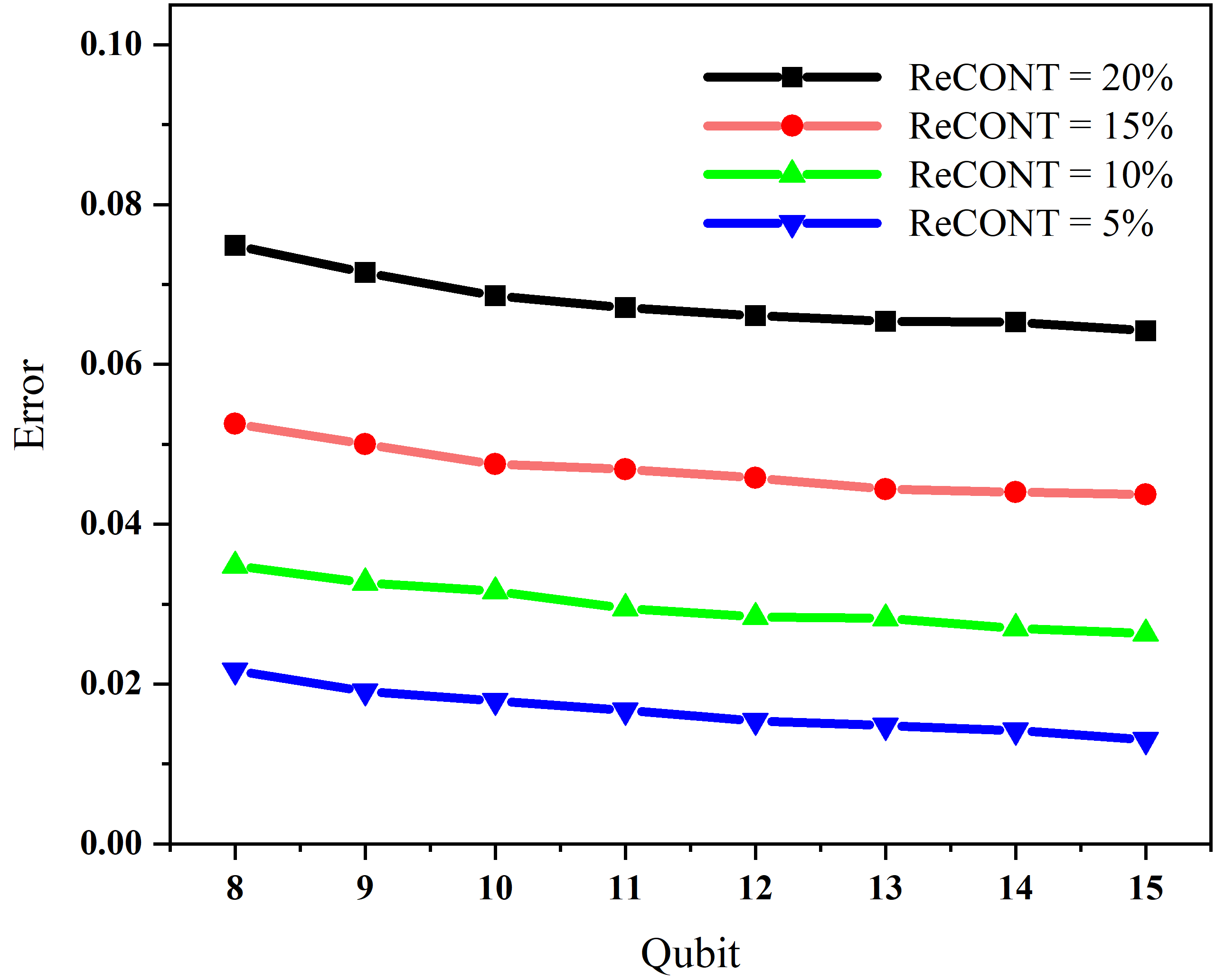}
\caption{The error values decrease with increasing qubit counts at a given CNOT gate reduction ratio}
\label{The error values decrease with increasing qubit counts at a given CNOT gate reduction ratios}
\end{figure}

\textbf{The performance of the algorithm improves with increasing qubit counts.} As shown in Figure \ref{The error values decrease with increasing qubit counts at a given CNOT gate reduction ratios}, under the same reduction in the proportion of CNOT gates, the error of the synthesized diagonal unitary decreases as the number of qubits increases. As shown in Table \ref{The error values and runtimes}, when the proportion of saved CNOT gates remains at 5\%, increasing the qubit counts from 8 to 15 reduces the error from 0.0217 to 0.0130. This can be attributed to the larger undirected graph constructed when more qubits are involved, which reduces the occurrence of ``dead-end''. The presence of ``dead-end'' leads to the selection of non-active nodes in the path, resulting in wasted CNOT gates. Therefore, under the same conditions for CNOT gate savings, fewer ``dead-end'' lead to lower errors in the synthesized diagonal unitary.

\begin{figure}[!h]
\centering
\includegraphics[scale=0.35]{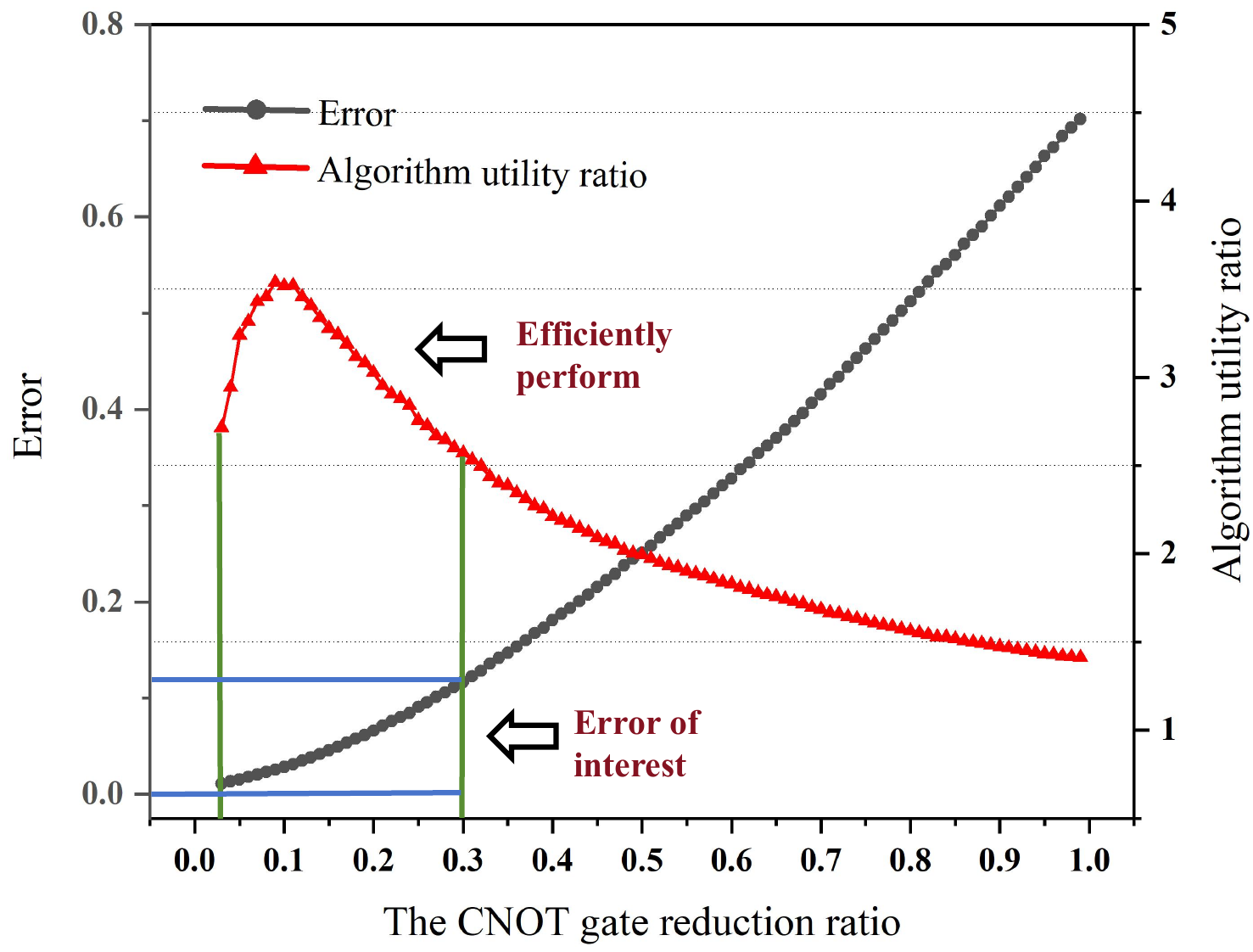}
\caption{The error values and the algorithm utility ratio obtained under various CNOT gate reduction ratios. Within the range of interest for algorithm error (0\%–12\%), the reduction ratio of CNOT gates remains below 30\%. In this range, the algorithm utility ratio exhibits outstanding performance.}
\label{The error values and the algorithm utility ratio obtained under various CNOT gate reduction ratios}
\end{figure}

\textbf{Within the error range of interest (0\%–12\%), the algorithm utility ratio outperforms other ranges.} As shown in Figure \ref{The error values and the algorithm utility ratio obtained under various CNOT gate reduction ratios}, when the number of qubits is 12 and the proportion of saved CNOT gates decreases from 30\% to 5\%, the algorithm utility ratio rises from 2.57 to 3.54. The reduction in the proportion of saved CNOT gates allows for more phase choices in the algorithm, leading to a greater number of selectable active nodes. As the number of selectable active nodes increases, the occurrence of ``dead-end'' decreases, which in turn reduces the chances of selecting non-active nodes along the path, thereby minimizing CNOT gate wastage. Moreover, an increase in the number of selectable active nodes also raises the probability of selecting high-quality active nodes. Consequently, the algorithm utility ratio improves.

\subsection{Runtime}
Our algorithm demonstrates impressive performance in terms of runtime, as shown in Table \ref{The error values and runtimes}. Within the error range of interest (0\%–12\%), the runtimes are as follows: when the number of qubits is 8, the algorithm runs for approximately 0.03 seconds; for 10 qubits, approximately 0.38 seconds; for 12 qubits, approximately 6.57 seconds; for 14 qubits, approximately 123.01 seconds; for 15 qubits, approximately 561.71 seconds. The algorithm's efficiency in terms of time is crucial for the synthesis and application of unitary.

Moreover, when the error in the synthesized diagonal unitary exceeds 25\%, the runtime required by our algorithm slightly increases. This is primarily because using too few CNOT gates increases the occurrence of ``dead-end'' situations, resulting in a higher computational overhead for node selection. However, since a substantial loss in algorithm fidelity is not practically meaningful, this error range is not our focus. Our algorithm still maintains excellent performance in terms of time within the error range of interest.

\section{CONCLUSION AND FUTURE WORK}
This paper explores the synthesis of diagonal unitary based on phase gadget techniques, revealing properties of phases and proposing an approximate algorithm for the synthesis of diagonal unitary. Within the error range of interest, the average utility ratio of the algorithm is 3.2. Furthermore, the algorithm performs well in terms of time efficiency; when the number of qubits is 15, the runtime of the algorithm within the error range of interest is approximately 561.71 seconds. Our algorithm achieves an improvement in circuit accuracy and a reduction in circuit cost at the expense of a certain level of algorithm precision loss, making it of significant practical relevance.

In the future, we aim to investigate whether a theoretical lower bound exists for the algorithm utility ratio and whether the correlation between the algorithm utility ratio and initial parameters can be quantified. While our work focuses on optimizing circuit size, further exploration is needed regarding the optimization of diagonal unitary in terms of circuit depth.


\newpage
\bibliographystyle{IEEEtran}
\bibliography{references}

\end{document}